\documentclass[twocolumn,twoside,slac_two]{revtex4}
\usepackage{graphicx}
\usepackage{fancyhdr}
\usepackage{multirow}

\input definitions

\begin{document}

\title{Measurements of CKM angles {\boldmath $\beta/\phi_1$} and
  {\boldmath $\alpha/\phi_2$} at the \babar\ and Belle experiments} 

%

\author{A.~Lazzaro on behalf of the \babar\ and Belle Collaborations}
\affiliation{Universit\`a degli Studi and INFN, Milano, Italy}

\begin{abstract}
We report measurements of the CKM angles $\beta/\phi_1$ and
$\alpha/\phi_2$ done by the \babar\ and Belle experiments.
Both experiments have collected large data samples, corresponding to
a total of more than 1 billion of \BB\ pairs, at the
$e^+e^-$ asymmetric-energy colliders PEP-II (SLAC) and KEK-B (KEK),
respectively.   
\end{abstract}

\maketitle

\thispagestyle{fancy}


\section{Introduction}

\CP\ violation in the Standard Model (SM)~\cite{SM} is described by an
irreducible complex phase in the Cabibbo-Kobayashi-Maskawa (CKM)
quark-mixing $3\times 3$  
matrix~\cite{CKM}. The equation $V_{ud}^{}V^\star_{ub} +
V_{cd}^{}V^\star_{cb}+V_{td}^{}V^\star_{tb}=0$, which follows from  
the unitarity of the CKM matrix $V$,
can be depicted as a triangle -- called the
Unitarity
Triangle (UT) -- in the complex plane~\cite{PDG08}. 
The main goal of the $B$-factories is to verify the SM picture of the
origin of the \CP\ violation by measuring the angles 
(denoted by $\alpha$, $\beta$, and $\gamma$~\footnote{Also denoted
by $\phi_2$, $\phi_1$, and $\phi_3$, respectively.
The Greek notation, used by \babar\ experiment, is used throughout this paper.})
and the sides of the UT in $B$ decays.
In this review we report results obtained by the \babar\ and Belle
collaborations concerning the measurements of the angles $\alpha$
and $\beta$.

\section{Detectors and Datasets}

Measurements reported in this paper have been obtained by the \babar\
and Belle experiments at the asymmetric-energy $e^+ e^-$ $B$ factories
PEP-II~\cite{pep2} and KEK-B~\cite{kekb}, respectively.
At the time of writing 
the two experiments collected more than $430\,\rm{fb^{-1}}$
and $750\,\rm{fb^{-1}}$, respectively, recorded at the \FourS\
resonance (center-of-mass energy $\sqrt{s}=10.58\ \gev$),
which corresponds to a total of approximately $1.3$ billion \BB\
events. 
PEP-II and \babar\ stopped 
data taking in April, 2008.
The \babar\ and Belle detectors are described
elsewhere~\cite{babardetector,belledetector}.

\section{Measurements of {\boldmath$\beta$}}
\label{sec:beta}

Measurements of time-dependent \CP\ asymmetries in $B^0$ meson decays
that proceed via the dominant CKM favored $b \rightarrow c
\bar{c} s$ tree amplitude, such as $\Bz \to J/\psi \Kz$,
have provided 
a precise measurement of angle $\beta$, giving
a crucial test of the
mechanism of \CP\ violation in the SM~\cite{sin2betaD}.
For such decays the interference between this amplitude and the
amplitude from $\Bz - \Bzb$ mixing is dominated by the single
phase $\beta = \arg{[-(V_{cd} 
V^*_{cb})/(V_{td} V^*_{tb})]}$ of the CKM mixing matrix.
Other quark transitions involving the charm quark which allow for the  
measurement of angle $\beta$, using time-dependent measurements of $B^0$
decays, are $b \to c\bar{c}d$ transitions, like $B^0 \to J/\psi \pi^0$ and $B^0 \to D^{*+}
D^{*-}$, and $b \to c\bar{u}d$ transitions,  like $B^0 \to D^{(*)0} h^0$.
Either tree and loop (penguin) amplitudes
can contribute in these transitions, so they are sensitive to New
Physics (NP) due to the large virtual mass scale occurring in the penguin loops.

To measure time-dependent \CP\ asymmetries we 
reconstruct a \Bz\ decaying into a
\CP\ eigenstate ($B_{\CP}$).  From the
remaining particles in the event we also reconstruct the decay
vertex of the other $B$ meson ($B_{\rm tag}$) and identify its flavor.
The difference $\deltat \equiv t_{\CP} - t_{\rm tag}$ of the proper decay
times $\tcp$ and $\ttag$ of the \CP\ and tag $B$ mesons, respectively, is
obtained from the measured distance between the $B_{\CP}$ and $B_{\rm
tag}$ decay vertices and from the known boost of the
\epem\ system.
The distribution of the difference
$\deltat$ is given by
\begin{eqnarray}
P(\dt) &=& 
 \frac{e^{-\left|\deltat\right|/\tau}}{4\tau} \{1 \pm 
 \label{eq:FCPdefPure}\\
 &&\hspace{-1em}
\left[-\eta_f S_f\sin(\deltamd\deltat) -
  C_f\cos(\deltamd\deltat)\right]\}\,\nonumber
\end{eqnarray}
where $\eta_f$ is the \CP\ eigenvalue of final state $f$,  the upper
(lower) sign denotes a decay accompanied by a \Bz (\Bzb) tag, $\tau$
is the mean \Bz\ lifetime, and $\deltamd$ is the mixing frequency. 
The parameters $C_f$ and $S_f$ for the final state $f$ are the
\CP-violating parameters 
\begin{equation}  
S_f = \frac{2 {\rm Im}(\lambda_f)}{|\lambda_f|^2 + 1}, \quad 
C_f = \frac{1-|\lambda_f|^2}{1+|\lambda_f|^2}, \nonumber
\end{equation} 
where $\lambda_f$ is a complex parameter depending on the $\Bz-\Bzb$
mixing as well as on the decay amplitudes 
for both \Bz\ and \Bzb\ to the \CP\ eigenstate~\footnote{Note that in
  the Belle convention $C_f = -A_f$.}. 

When only one diagram contributes to the
decay process and no other weak or strong phases
appear in the process, the SM predicts $C_f = 0$ and $S_f = - \eta_f
\sin 2 \beta$. A nonzero value of the parameter $C_f$ would indicate
direct \CP\ violation.  Any significant deviation from the SM
prediction could be a sign of NP.

An alternative way to measure the angle $\beta$ is to use
measurements of time-dependent \CP\ asymmetries in
decays of $B^0$ mesons to charmless hadronic final states, such as
\phikz, \fzkz, \kpkmkz, \etaprkz, \pizkzs, \kzskzskzs, \rhozkzs,
and \omegakzs.
These decays are CKM-suppressed $b\to q\bar{q}s$ ($q=u,\, d, \, s$)
processes that are dominated by a single penguin amplitude, with the
same weak phase as the $b \to c \bar{c} s$
transition~\cite{Penguin}. In these modes, assuming the penguin
dominance of the \btos\ transition and neglecting CKM-suppressed amplitudes, the
time-dependent 
\CP-violating parameter 
$S_f$ is expected to be $-\eta_f \stwob$.
However, CKM-suppressed amplitudes and the color-suppressed
tree-level diagram introduce additional 
weak phases whose contribution may not be
negligible~\cite{london,Gross,Gronau,BN}.
As a consequence, only an effective $S = -\eta_f \stwob_{\rm eff}$
is determined. The deviation $\Delta S = S - (-\eta_f \stwob)$
has been estimated in
several theoretical approaches~\cite{Gross,Gronau,BN,BN2,Cheng,Zupan}. 
The estimates are channel and model dependent.
Also for these decays
the possible presence of additional diagrams with new heavy particles in
the loop and new
\CP-violating phases may contribute to the decay amplitudes.
In this case the measurements of significantly larger $\Delta S$ are a
sensitive probe for NP~\cite{london}.

\subsection{\boldmath $b \to c\bar{c}s$ Decays}

Decays underlain by $b \to c\bar{c}s$ transitions are referred to as
``golden modes'' 
due to their relatively large branching fractions
$\mathcal{O}(10^{-4}-10^{-5})$, low 
experimental background levels and high reconstruction efficiencies.
They are dominated by a color-suppressed tree diagram and 
the theoretical uncertainties are small~\cite{golden}.
Hence the prediction $S_f = -\eta_f \sin 2 \beta$ and $C_f = 0$ is a
good approximation.

\babar\ reconstructed the modes $B^0$ to $J/\psi \KS$,
$J/\psi K^{*0}$, $\psi(2S) \KS$, $J/\psi \KL$, 
 $\eta_{c} \KS$, and $\chi_{c1} \KS$,
extracting the \CP-violating parameters from a simultaneous fit to all modes.
The amount of data used corresponds to $383$ million \BB\
pairs~\cite{ccsBabar}. 
Belle reconstructed the modes 
$B^0 \to J/\psi \KS$ and $B^0 \to J/\psi \KL$ 
using 535 million \BB\ pairs~\cite{ccsBelle}.
Recently Belle also reported a measurement of the
\CP-violating parameters in the $B^0 \to \psi(2S) \KS$
channel, using a sample of 657 million \BB\ pairs~\cite{psiksBelle}.
The \babar\ and Belle results agree within the measurement uncertainties.   
All results are shown in Table~\ref{tab:golden}.
A world average,
calculated by the Heavy Flavor Averaging Group (HFAG)~\cite{HFAG}, 
gives 
$\sin 2\beta = 0.680 \pm 0.025$,
which reduces the total uncertainty on $\sin 2\beta$ to $3.7\%$.  
No evidence of direct \CP\ violation is seen in these modes.

\begin{table}[h]
\begin{center}
\caption{Results for decays involving $b\to c\bar{c}s$ transitions(``golden modes'').
The errors are, in order, statistical and systematic.}
\begin{tabular}{l|c|c}
\hline \hline
\multicolumn{1}{c|}{\multirow{2}{*}{$B^0$ decays}} & \BB\  pairs&
\multirow{2}{*}{Results} \\ 
 & ($\times 10^6$) & \\

\hline \hline
\multicolumn{3}{c}{\babar} \\ \hline
$J/\psi \KS$, & \multirow{6}{*}{383} &
\multirow{6}{4.7cm}{
$\sin 2\beta  = \phantom{+}0.714 \pm 0.032 \pm 0.018$
$\phantom{\sin 2}\,C = \phantom{+}0.049 \pm 0.022 \pm 0.017$} \\
$J/\psi K^{*0}$, & \\
$\psi(2S) \KS$, & \\ 
$J/\psi \KL$, & \\
$\eta_{c} \KS$, & \\
$\chi_{c1} \KS$ & \\
\hline \hline
\multicolumn{3}{c}{Belle} \\ \hline
$J/\psi \KS$, & \multirow{2}{*}{535} & 
\multirow{2}{4.7cm}{
$\sin 2\beta  = \phantom{+}0.642 \pm 0.031 \pm 0.017$
$\phantom{\sin 2}\,\,C = -0.018 \pm 0.021 \pm 0.014$} \\
$J/\psi \KL$ &  \\ \hline
\multirow{2}{*}{$\psi(2S) \KS$} & \multirow{2}{*}{657} & 
\multirow{2}{4.7cm}{
$\sin 2\beta  = \phantom{+}0.718 \pm 0.090 \pm 0.033$
$\phantom{\sin 2}\,C = -0.039 \pm 0.069 \pm 0.049$} \\
 &  \\ \hline
\multirow{2}{*}{Average} & \multirow{2}{*}{$-$} & 
\multirow{2}{4.7cm}{
$\sin 2\beta  = \phantom{+}0.650 \pm 0.029 \pm 0.018$
$\phantom{\sin 2}\,C = -0.019 \pm 0.020 \pm 0.015$} \\
 &  \\ 
\hline \hline
\end{tabular}
\label{tab:golden}
\end{center}
\end{table}

\babar\ and Belle have also reported measurements of the $\beta$ angle using
the \Bz\ decay to
$D^{*\pm}D^{*\mp}\KS$~\cite{dsdskBabar,dsdskBelle}. This decay
proceeds mainly through the 
$b\to c\bar{c}s$ transition. 
A potential interference effect of the
decay proceeding through an intermediate resonance can be measured by
dividing the $B$-decay Dalitz plot into regions with 
\mbox{$s^+ \le s^-$} or \mbox{$s^+ \ge s^-$}, where $s^{\pm} \equiv
m^2 (D^{*\pm} K_{\rm{S}}^0)$.
Such an analysis offers the interesting possibility to
extract the sign of $\cos 2\beta$, therefore partially resolving
the 4-fold ambiguity in the 
value of $\beta$ obtained from the measurement of the $\sin 2\beta$.
For these modes the time-dependent \CP\ asymmetry is described
in terms of the 
coefficients $J_c$, $J_0$, $J_{s1}$, and $J_{s2}$, 
which are the integrals over the half-Dalitz space.
The results are shown in Table~\ref{tab:DsDsKs} and there is a general
agreement between the two experiments.
\babar\ infers that $\cos 2\beta>0$ at 94\% confidence level (CL),
on the assumption that $J_{s2}>0$~\cite{dsdskBabar}.

\begin{table}[h]
\begin{center}
\caption{Results for \Bz\ decay to $D^{*\pm}D^{*\mp}\KS$.
The errors are, in order, statistical and systematic.}
\begin{tabular}{l|c|c}
\hline \hline
 & \BB\  pairs&
\multirow{2}{*}{Results} \\ 
 & ($\times 10^6$) & \\
\hline \hline
\multirow{3}{*}{\babar} & \multirow{3}{*}{230} & 
\multirow{3}{5.1cm}{
$\phantom{2\sin 2\beta} J_c/J_0  = \phantom{+}0.76 \pm 0.18 \pm 0.07$
$2J_{s1}/J_0 \sin 2\beta = \phantom{+}0.10 \pm 0.24 \pm 0.06$ 
$2J_{s2}/J_0 \cos 2\beta = \phantom{+}0.38 \pm 0.24 \pm 0.05$} \\
& \\
&\\
\hline
\multirow{3}{*}{Belle} & \multirow{3}{*}{449} &
\multirow{3}{5.1cm}{
$\phantom{2\sin 2\beta} J_c/J_0  = \phantom{+}0.60^{+0.25}_{-0.28} \pm 0.08$
$2J_{s1}/J_0 \sin 2\beta = -0.17 \pm 0.42 \pm 0.09$ 
$2J_{s2}/J_0 \cos 2\beta = -0.23^{+0.43}_{-0.41} \pm 0.13$} \\
& \\
&\\
\hline\hline
\end{tabular}
\label{tab:DsDsKs}
\end{center}
\end{table}

\subsection{\boldmath $b \to c\bar{c}d$ Decays}

The $B^0 \to J/\psi \piz$ decay takes place through a \mbox{$b \to
  c\bar{c}d$} 
transition.  The dominant tree diagram is Cabibbo suppressed.
However there is a penguin diagram of
the same order as the tree diagram but with a different weak phase.
So, contrary to the golden modes,  even within the SM, the deviation
of $S$ measured in $b \to
  c\bar{c}d$ modes
from $-\eta_f \sin 2\beta$ could be substantial. 
Both \babar\ and Belle have updated measurements for this mode, which
are shown in Table~\ref{tab:ccdModes}~\cite{Jpsipi0Babar,Jpsipi0Belle}.
In particular the \babar\ result provides evidence of \CP\ violation,
with a statistical significance
of $4\sigma$, while for Belle it is $2.4\sigma$.

The decay $B^0 \to D^{*+} D^{*-}$ also goes through the $b \to
c\bar{c}d$ transition.
This mode requires an angular analysis to disentangle 
\CP-odd and \CP-even events.
Results for \CP-violating parameters are show in
Table~\ref{tab:ccdModes}~\cite{dsdsBabar}. 

The quark transition $b\to c\bar{c}d$ is also responsible for the
\Bz\ decays to $D^{*+}D^-$, $D^{*-}D^+$, and $D^+D^-$.
Results for the \babar\ and Belle experiments are shown in
Table~\ref{tab:ccdModes}~\cite{ddBabar,ddBelle}.
Belle reports a statistical significance of $3.2\sigma$ for direct \CP\ violation in the
$D^+D^-$ mode, while \babar\ reports $0.4\sigma$.

Within the experimental uncertainties, all results for $b\to c\bar{c}d$
decays are compatible with the SM prediction.

\begin{table}[h]
\begin{center}
\caption{Results for \Bz\ decays to $J/\psi \piz$, $D^{*+}D^{*-}$,
$D^{*+}D^-$, $D^{*-}D^+$, and $D^+D^-$ ($b \to c\bar{c}d$ decays).
The errors are, in order, statistical and systematic.}
\begin{tabular}{l|c|c}
\hline \hline
 & \BB\  pairs&
\multirow{2}{*}{Results} \\ 
 & ($\times 10^6$) & \\
\hline \hline
\multicolumn{3}{c}{$J/\psi \piz$} \\
\hline
\multirow{2}{*}{\babar} & \multirow{2}{*}{466} & 
$S = -1.23 \pm 0.21 \pm 0.04$ \\
& & $C = -0.20 \pm 0.19 \pm 0.03$ \\
\hline
\multirow{2}{*}{Belle} & \multirow{2}{*}{535} & 
$S = -0.65 \pm 0.21 \pm 0.05$ \\
& & $C = -0.08 \pm 0.16 \pm 0.05$ \\
\hline\hline
\multicolumn{3}{c}{$D^{*+}D^{*-}$} \\ \hline
\hline
\multirow{2}{*}{\babar} & \multirow{2}{*}{383} & 
$S = -0.66 \pm 0.19 \pm 0.04$ \\
& & $C = -0.02 \pm 0.11 \pm 0.02$ \\
\hline
\multirow{2}{*}{Belle} & \multirow{2}{*}{657} & 
$S = -0.93 \pm 0.24 \pm 0.15$ \\
& & $C = -0.16 \pm 0.13 \pm 0.02$ \\
\hline \hline
\multicolumn{3}{c}{$D^{*+}D^-$} \\ \hline
\multirow{2}{*}{\babar} & \multirow{2}{*}{383} & 
$S = -0.79 \pm 0.21 \pm 0.06$ \\
& & $C = \phantom{+}0.18 \pm 0.15 \pm 0.04$ \\
\hline
\multirow{2}{*}{Belle} & \multirow{2}{*}{152} & 
$S = -0.55 \pm 0.39 \pm 0.12$ \\
& & $C = -0.37 \pm 0.22 \pm 0.06$ \\
\hline \hline
\multicolumn{3}{c}{$D^{*-}D^+$} \\ \hline
\multirow{2}{*}{\babar} & \multirow{2}{*}{383} & 
$S = -0.44 \pm 0.22 \pm 0.06$ \\
& & $C = \phantom{+}0.23 \pm 0.15 \pm 0.04$ \\
\hline
\multirow{2}{*}{Belle} & \multirow{2}{*}{152} & 
$S = -0.96 \pm 0.43 \pm 0.12$ \\
& & $C = \phantom{+}0.23 \pm 0.25 \pm 0.06$ \\
\hline \hline
\multicolumn{3}{c}{$D^+D^-$} \\ \hline
\multirow{2}{*}{\babar} & \multirow{2}{*}{383} & 
$S = -0.54 \pm 0.34 \pm 0.06$ \\
& & $C = \phantom{+}0.11 \pm 0.22 \pm 0.07$ \\
\hline
\multirow{2}{*}{Belle} & \multirow{2}{*}{535} & 
$S = -1.13 \pm 0.37 \pm 0.09$ \\
& & $C = -0.91 \pm 0.23 \pm 0.06$ \\
\hline\hline
\end{tabular}
\label{tab:ccdModes}
\end{center}
\end{table}

\subsection{\boldmath $b \to c\bar{u}d$ Decays}

The decay $B^0 \to D_{\CP}^{(*)0}  h^0 (h^0 = \pi^0, \eta, \omega)$ is
governed by a color-suppressed $b \to c\overline{u}d$ tree diagram. 
When the
neutral $D$ meson decays to a \CP\ eigenstate Eq.~\ref{eq:FCPdefPure}
is still valid.
In these modes the possible effects of NP are expected to be small, so we
expect $S = \sin2\beta$~\cite{Penguin}.
Only \babar\ reported a measurement of such decays~\cite{dcph0}, by
reconstructing the following decay modes  $D^{*0} \to D^0 \pi^0$ and
$D^0 \to K^+ K^-$, 
$D^0 \to K_{\rm{S}}^0 \pi^0$ and $D^0 \to K_{\rm{S}}^0 \omega$.  The
analysis is performed using $383 \times
  10^6\,B\overline{B}$ pairs from which $340 \pm 32$ signal events are
  reconstructed.  The measured \CP-violating parameters,
\begin{eqnarray}
\sin 2\beta&=&  \phantom{+}0.56 \pm 0.23{\rm (stat)} \pm 0.05{\rm
  (syst)} \nonumber \\ 
C &=& -0.23 \pm 0.16{\rm (stat)} \pm 0.04{\rm (syst)} \nonumber, 
\end{eqnarray}
are consistent with the SM expectations. 

Also the decay $B^0 \to D^{(*)0} h^0$, where $h^0 = \piz, \eta, \omega,
\etapr$, is governed by the $b \to c\overline{u}d$ tree diagram. 
This decay can occur with and without $\Bz-\Bzb$
mixing and interference effects being visible across the $D^0 \to \KS
\pi^+ \pi^-$ Dalitz plot. 
The interesting result from this measurement is the possibility to
extract the sign of  $\cos 2 \beta$ in order to resolve
the 4-fold ambiguity in the 
value of $\beta$ obtained from the measurement of the $\sin 2\beta$.
The results from \babar\ are obtained using 
$383$ million of \BB\ pairs:
\begin{eqnarray}
\sin 2\beta&=&  0.29 \pm 0.34 {\rm (stat)} \pm 0.03{\rm (syst)} \pm 0.05{\rm (Dalitz)} \nonumber \\
\cos 2\beta&=& 0.42 \pm 0.49{\rm (stat)} \pm 0.09{\rm (syst)} \pm 0.13{\rm (Dalitz)} \nonumber, 
\end{eqnarray}
leading to a preferred positive sign for $\cos 2\beta$ at
$86 \%$ CL~\cite{dalitzdhBabar}.
The Dalitz error refers to the Dalitz model parameterization used
in the analysis.
Belle performed a similar analysis 
using 386 million of \BB\ pairs:
\begin{eqnarray}
\sin 2\beta&=&  0.78 \pm 0.44{\rm (stat)} \pm 0.22{\rm (syst + Dalitz)} \nonumber \\
\cos 2\beta&=& 1.87^{+0.40}_{-0.53}{\rm (stat)}^{+0.22}_{-0.32}{\rm (syst + Dalitz)} \nonumber, 
\end{eqnarray}
which gives a preferred positive sign of $\cos 2\beta$ at
$98.3 \%$ CL~\cite{dalitzdhBelle}.

\subsection{\boldmath $b \to s$ Decays}

\begin{figure}[h]
\centering
\includegraphics[width=80mm]{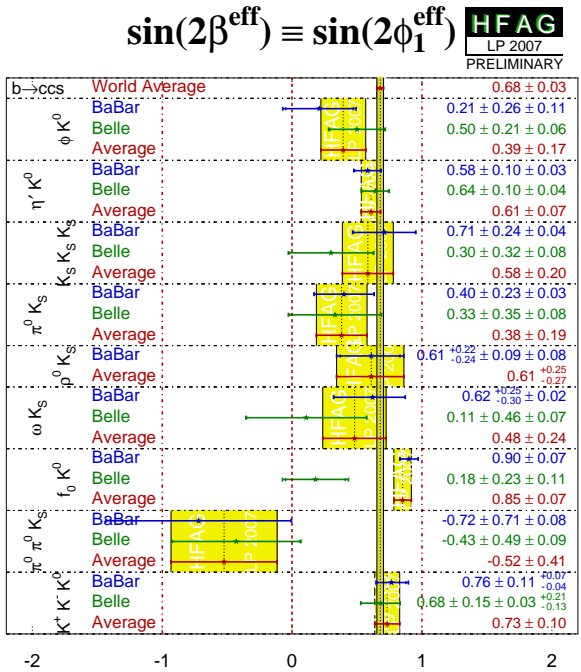}
\caption{Summary of time-dependent $S$ parameter results for $b\to s$
  penguin modes~\cite{HFAG}. } 
\label{fig:btos}
\end{figure}

No major updates on the $b \to s$ decays have been reported by the
\babar\ and Belle recently. The summary of results for the
time-dependent $S$ parameter is shown in
Fig.~\ref{fig:btos}~\cite{HFAG}.
In general the results are consistent between \babar\ and Belle, and
consistent with SM expectations within the statistical uncertainties. 
Some tensions are observed for 
the $\Bz \to \piz\piz\KS$ decay results 
with respect to the SM expectation.
Also tensions are observed in $\Bz \to f_0(980) \KS$ decay between
\babar\ and Belle results. However, in this case the \babar\ result
reported in the figure is a combination of results from the two Dalitz
plot analyses, considering $f_0(980) \to K^+K^-$ and $f_0(980) \to
\pi^+\pi^-$, while Belle uses only the $f_0(980) \to \pi^+\pi^-$ mode.
The results are found to be in agreement if only the $f_0(980) \to
\pi^+\pi^-$ decay is considered. 
No evidence of direct \CP\ violation is observed.

\section{Measurements of {\boldmath$\alpha$}}

The UT angle $\alpha$, defined as 
$\arg{[-(V_{td}^{}V_{tb}^*)/(V_{ud}^{}V_{ub}^*)]}$, can be determined by 
measuring a  
time-dependent \CP\ asymmetry in charmless $b\rightarrow
u\overline{u}d$ decays  
such as $B^0\rightarrow\pi^+\pi^-,\,\pi^+\pi^-\pi^0,\,\rho^+\rho^-$, and 
$a_1^\pm(1260)\pi^\mp$, in a way similar to what described in 
section~\ref{sec:beta}.
The $B$ decays proceed
mainly through a tree and gluonic penguin amplitude.

\subsection{{\boldmath$B^0 \to \pip\pim$} and {\boldmath$B^0 \to
    \rho^+\rho^-$} Decays}

Similar to Eq.~\ref{eq:FCPdefPure}, the time-dependent rate for $\Bz
\to \pi^+\pi^-$ is given by
\begin{eqnarray}
P(\Delta t) =  \frac{e^{-|\Delta
    t|/\tau}}{4\tau}\{ 1  \pm
 \label{eq:FCPPiPi}\\
 &&\hspace{-6em}
\left[ S\sin(\deltamd\deltat) -
  C \cos(\deltamd\deltat)\right]\}\,\nonumber
\end{eqnarray}
where
$C$  and $S$  
are \CP\ asymmetry coefficients.
If the decay amplitude is dominated by a tree diagram then
$S = \sin 2\alpha$ and $C = 0$. The presence 
of an amplitude with a different weak phase (such as from a gluonic
penguin diagram) gives rise  
to direct \CP\ violation and shifts $S$ from $\sin 2\alpha$ to 
\begin{eqnarray}
S = \sqrt{1 - C^2} \sin 2\alpha_{\rm eff},
\label{eq:phi2eff}
\end{eqnarray} 
where $\alpha_{\rm eff} = \alpha + \delta \alpha$, and $\delta \alpha$ is the
phase shift.  

The \babar\ and Belle results for $\Bz \to \pip\pim$ time-dependent
\CP-violating parameters are shown in
Table~\ref{tab:AlphaModes}~\cite{PiPiBabar, PiPiBelle}. 
Both measurements indicate a large mixing-induced \CP-violation 
with a significance greater than $5\sigma$ independent of the value of $C$.
Belle has also observed large direct \CP\ violation
($5.5\sigma$). 
The difference between \babar\ and Belle on the $(S, \,C)$ plane 
is about $2.1\sigma$~\cite{HFAG}.
The angle $\alpha$ can be extracted using isospin relations, which
require the measurement of the branching fractions and \CP-violating
parameters 
for the SU(2) partners of the $B^0 \to \pip\pim$
decay: $B^0 \to \piz\piz$ and $B^\pm \to \piz\pi^\pm$~\cite{GL}.  
The \babar\ and Belle constraints on $\alpha$
are consistent with the SM and are given by $(96^{+10}_{-6})^\circ$ 
and $(97\pm 11)^\circ$, respectively, at $68\%$ CL.

Another decay used to measure the angle $\alpha$ is $B^0 \to
\rho^+\rho^-$. 
For this mode the same considerations described above for
the $\Bz \to \pip \pim$ mode are still valid. 
The \babar\ and Belle results are shown in
Table~\ref{tab:AlphaModes}~\cite{RhoRhoBabar, RhoRhoBelle}.
Both experiments are consistent with each other
and consistent with no \CP\ violation. 
To extract $\alpha$ using SU(2) relations, 
\babar\ has recently performed the measurement of the time-dependent
\CP-violating parameters in $B^0 \to \rho^0\rho^0$ with 427 million of \BB\
pairs, measuring a 
branching fraction of 
$(0.84\pm0.29\pm0.17) \times 10^{-6}$, a longitudinal polarization of
$0.70 \pm 0.14 \pm 0.05$  and for the longitudinally polarized events
$S_L=0.5\pm 0.9 \pm 0.2$ and $C_L=0.4\pm 0.9\pm 0.2$ (errors are
statistical and systematic, respectively)~\cite{BabarRho0Rho0}.
Together with all other measurements from SU(2) partners, it
results $74^\circ < \alpha < 117^\circ$
at 68\% CL, with a constraint of 
$|\delta \alpha|<14.5^\circ$ at 68\% CL and a preferred
solution of $\delta \alpha = +11.3^\circ$. Belle, using data of 657
million of \BB\ pairs, set an upper limit on
the branching fraction for $B^0 \to \rho^0\rho^0$ of $1.0\times
10^{-6}$ at 90\% CL, and have yet to measure the time-dependent
\CP-violating parameters for this mode.
The Belle constraint on $\alpha$ is $(91.7\pm14.9)^\circ$~\cite{BelleRho0Rho0}.

\begin{table}[h]
\begin{center}
\caption{Results for \Bz\ decays to $\pi^+ \pi^-$ and
$\rho^+\rho^-$.
Note that the \CP-violating parameters for $\Bz \to \rho^+\rho^-$ refer to
longitudinally polarized events.
The errors are, in order, statistical and systematic.}
\begin{tabular}{l|c|c}
\hline \hline
 & \BB\  pairs&
\multirow{2}{*}{Results} \\ 
 & ($\times 10^6$) & \\
\hline \hline
\multicolumn{3}{c}{$\pip \pim$} \\
\hline
\multirow{2}{*}{\babar} & \multirow{2}{*}{383} & 
$S = -0.60 \pm 0.11 \pm 0.03$ \\
& & $C = -0.21 \pm 0.09 \pm 0.02$ \\
\hline
\multirow{2}{*}{Belle} & \multirow{2}{*}{535} & 
$S = -0.61 \pm 0.10 \pm 0.04$ \\
& & $C = -0.55 \pm 0.08 \pm 0.05$ \\
\hline\hline
\multicolumn{3}{c}{$\rho^+ \rho^-$} \\ \hline
\hline
\multirow{2}{*}{\babar} & \multirow{2}{*}{383} & 
$S_L = -0.17 \pm 0.20^{+0.05}_{-0.06}$ \\
& & $C_L = \phantom{+}0.01 \pm 0.15 \pm 0.06$ \\
\hline
\multirow{2}{*}{Belle} & \multirow{2}{*}{535} & 
$S_L = \phantom{+}0.19 \pm 0.30 \pm 0.08$ \\
& & $C_L = -0.16 \pm 0.21 \pm 0.08$ \\
\hline \hline
\end{tabular}
\label{tab:AlphaModes}
\end{center}
\end{table}

\subsection{{\boldmath$B^0 \to \pip\pim\piz \,\, (\rho\pi)^0$} and
  {\boldmath$B^0 \to a_1^\pm(1260)\pi^\mp$} Decays}

An alternative way to measure the angle $\alpha$ is to perform 
a time-dependent Dalitz plot analysis in $B^0\rightarrow \pi^+\pi^-\pi^0$
decays. We model the interference between the intersecting $\rho$
resonance bands and so determines the strong phase differences from the
Dalitz plot structure~\cite{snyder_quinn}.
The Dalitz amplitudes and time-dependence are contained in complex
parameters that are determined by fit on data. 
This technique allows to extract directly $\alpha$.
\babar\ and Belle have performed measurements using 383 million and
449 million of \BB\ pairs, respectively. 
The intervals at 68\% CL are $74^\circ < \alpha <
132^\circ$ for \babar~\cite{rhopiBabar} and $68^\circ < \alpha <
95^\circ$ for Belle~\cite{rhopiBelle}.

Another channel which allows for a measurement of $\alpha$ is 
$B^0 \rightarrow a_1^\pm(1260) \pi^\mp$. 
For this mode a Dalitz plot analysis is not feasible with current
statistics, so a quasi-two-body approach is used.
As the final state $a_1^\pm(1260)\pi^\mp$ 
is not a \CP\ eigenstate, one has to consider four decay modes,
divided in two groups, with
different charge and flavor combinations: 
$\Bz \to a_1^+(1260) \pi^-$ and $\Bzb \to a_1^+(1260) \pi^-$;
$\Bz \to a_1^-(1260) \pi^+$ and $\Bzb \to a_1^-(1260) \pi^+$.
Equation~\ref{eq:FCPPiPi} is valid for each group, 
where we denote the \CP-violating parameters as 
$S^+$, $C^+$ and $S^-$, $C^-$, respectively~\cite{a1pi}. 
It is possible to redefine these parameters as $S = (S^+ + S^-)/2$, $C
= (C^+ + C^-)/2$,  $\Delta S = (S^+ - S^-)/2$, $\Delta C = (C^+ -
C^-)/2$. \babar\ performed this analysis using 383 million of \BB\
pairs~\cite{a1pibabar}, extracting $608\pm52$ signal events and the
following time-dependent \CP-violating parameters:
\begin{eqnarray}
S&=&  \phantom{+}0.37 \pm 0.21 \pm 0.07 \nonumber \\
C &=& -0.10 \pm 0.15 \pm 0.09 \nonumber \\
\Delta S &=& -0.14 \pm 0.21 \pm 0.06 \nonumber \\
\Delta C &=& \phantom{+}0.26 \pm 0.15 \pm 0.07 \nonumber
\end{eqnarray}
where the errors are, in order, statistical and systematic.
Also a time- and flavour-integrated charge asymmetry for direct
\CP\ violation has been measured, ${\cal A}_{\CP} = -0.07\pm0.07\pm0.02$.
These measurements indicate no
direct and time-dependent \CP\ violation in $B^0\to a_1^\pm(1260)
\pi^\mp$ decay. 
The effective angle $\alpha_{\rm eff}$ is $(78.6 \pm 7.3)^\circ$.
The extraction of $\alpha$ can be performed by evoking SU(3) flavor
symmetry~\cite{a1pi}.  
Once the measurements of the branching fractions for the 
SU(3)-related decays become available, 
it will be possible to determine 
an upper bound on $\delta
\alpha$ in $B^0 \rightarrow a_1^\pm(1260) \pi^\mp$ decays and therefore yield
a constraint on the angle $\alpha$.

\section{Conclusions}
In this review we have presented measurements 
done by the \babar\ and Belle experiments that are 
used to determine
the angles $\beta$ and $\alpha$ of the UT.
The world averages values are $\beta = (21.5 \pm
1.0)^\circ$~\cite{HFAG} and $\alpha = (87.5^{+6.2}_{-5.3})^\circ$~\cite{CKMfit}.
The \CP-violating parameters are consistent with the
Standard Model expectations within the uncertainties of the
measurements. 

\begin{acknowledgments}
I would like to thank all my \babar\ collaborators, in particular
V.~Lombardo and F.~Palombo,
the conference organization and the Australian people for their kind
hospitality. 
\end{acknowledgments}

\bigskip 

\end{document}